\documentclass[preprint,showpacs,preprintnumbers,amsmath,amssymb]{revtex4}
\usepackage{epsfig}
\usepackage{graphicx}
\usepackage{dcolumn}
\usepackage{bm}

\newcommand{\ord}{{\cal O}}
\def\beq{\begin{equation}}
\def\eeq{\end{equation}}
\def\eeqn{\end{equation}}
\newcommand\iden{\leavevmode\hbox{\small1\normalsize\kern-.33em1}}

\newcommand{\sq}{\sqrt{2}}
\newcommand{\bea} {\begin{eqnarray}}
\newcommand{\eea} {\end{eqnarray}}

\newcommand{\Gm}{\Gamma}
\newcommand{\sbt}{s_{\beta}}
\newcommand{\cbt}{c_{\beta}}
\newcommand{\tbt}{t_{\beta}}

\let\jnfont=\rm
\def\NPB#1,{{\jnfont Nucl.\ Phys.\ B }{\bf #1},}
\def\PLB#1,{{\jnfont Phys.\ Lett.\ B }{\bf #1},}
\def\EPJC#1,{{\jnfont Eur.\ Phys.\ Jour.\ C }{\bf #1},}
\def\PRD#1,{{\jnfont Phys.\ Rev.\ D }{\bf #1},}
\def\PRL#1,{{\jnfont Phys.\ Rev.\ Lett.\ }{\bf #1},}
\def\MPLA#1,{{\jnfont Mod.\ Phys.\ Lett.\ A }{\bf #1},}
\def\JPG#1,{{\jnfont J.\ Phys.\ G }{\bf #1},}
\def\CTP#1,{{\jnfont Commun.\ Theor.\ Phys.\ }{\bf #1},}
\def\JHEP#1,{{\jnfont JHEP \ }{\bf #1},}
\def\NPPS#1,{{\jnfont Nucl.\ Phys.\ Proc.\ Suppl.\ }{\bf #1},}
\def\CPC#1,{{\jnfont Computl.\ Phys.\ Commun.\ }{\bf #1},}
\def\CPL#1,{{\jnfont Chin.\ Phys.\ Lett. }{\bf #1},}
\def\APPB#1,{{\jnfont Acta\ Phys.\ Polon.\ B }{\bf #1},}

\def\lsim{\raise0.3ex\hbox{$<$\kern-0.75em\raise-1.1ex\hbox{$\sim$}}}
\def\gsim{\raise0.3ex\hbox{$>$\kern-0.75em\raise-1.1ex\hbox{$\sim$}}}

\begin{document}

\title{\ \\[10mm] The LHC di-photon Higgs signal
                  predicted by little Higgs models }

\author{Lei Wang$^1$, Jin Min Yang$^{2}$}

\affiliation{
$^1$ Department of Physics, Yantai University, Yantai 264005, PR China\\
$^2$ Key Laboratory of Frontiers in Theoretical Physics,\\
     Institute of Theoretical Physics, Academia Sinica,
             Beijing 100190, PR China \vspace{0.5cm} }


\begin{abstract}
Little Higgs theory naturally predicts a light Higgs boson whose
most important discovery channel at the LHC is the di-photon signal
$pp\to h\to \gamma\gamma$. In this work we perform a comparative
study for this signal in some typical little Higgs models, namely
the littlest Higgs model (LH), two littlest Higgs models with
T-parity (named LHT-I and LHT-II) and the simplest little Higgs
modes (SLH). We find that compared with the Standard Model
prediction, the di-photon signal rate is always suppressed and the
suppression extent can be quite different for different models. The
suppression is mild ($\lsim 10\%$) in the LH model but can be quite
severe ($\simeq 90\%$) in other three models. This means that
discovering the light Higgs boson predicted by the little Higgs
theory through the di-photon channel at the LHC will be more
difficult than discovering the SM Higgs boson.
\end{abstract}


\maketitle

\section{Introduction}
The little Higgs \cite{LH} is proposed as an elegant mechanism
of electroweak symmetry breaking with a naturally light Higgs sector.
So far various realizations of the little Higgs symmetry structure
have been proposed \cite{otherlh,lst,sst}, which can be categorized
generally into two classes \cite{smoking}. One class use the product
group, represented by the littlest Higgs model (LH) \cite{lst}, in
which the SM $SU(2)_L$ gauge group is from the diagonal breaking of
two (or more) gauge groups. The other class use the simple group,
represented by the simplest little Higgs model (SLH) \cite{sst}, in
which a single larger gauge group is broken down to the SM
$SU(2)_L$. Further, to relax the constraints from the electroweak
precision tests \cite{sst,cstrnotparity}, a discrete symmetry called
T-parity is proposed \cite{tparity}, which can also provide a candidate
for the cosmic dark matter.
For the LH there are two different implementations of
T-parity in the fermion sector, called respectively LHT-I and
LHT-II \cite{lhti,lhtii}. A characteristic difference between
LHT-I and LHT-II is that the top quark partner responsible for
canceling the one-loop quadratic divergence of Higgs mass
contributed by the top quark is T-even for the former
and T-odd for the latter. The implementation of T-parity
in the SLH has also been tried \cite{slmtparity}.

To test the little Higgs theory at the LHC, the Higgs phenomenology
will play an important role \cite{lh-higgs-lhc}. At the LHC
different search strategies will be applied for different mass
ranges. For a light Higgs boson below about 140 GeV the di-photon
signal $pp\to h\to \gamma\gamma$ is the most important discovery
channel because the narrow $\gamma\gamma$ peak can be reconstructed
to distinguish the signal from the backgrounds. In contrast, the
dominant channel $pp\to h\to \bar bb$ cannot be utilized for
discovery because of the overwhelming QCD backgrounds. Recently the
ATLAS collaboration reported their di-photon search results with 209
$pb^{-1}$ of data collected early 2011 and excluded a signal rate of
4.2-15.8 times the SM prediction for 110 GeV $\leq m_h\leq$ 140 GeV
\cite{newhggrr}. With a luminosity of 2 $fb^{-1}$ the ongoing LHC
will be able to use the di-photon signal to exclude a light SM Higgs
boson. So the di-photon Higgs channel will be a sensitive probe for
new physics models like the little Higgs theory.

So far the di-photon signal has been studied in some new physics
models \cite{nph,hrrhan,lhtiyuan,higgslhtii}. Although some little
Higgs models have also been discussed
\cite{hrrhan,lhtiyuan,higgslhtii}, these previous studies are
performed separately in different frameworks. To show the difference
of model predictions, it is necessary to perform a comparative study
for various models. Further, the study for the SLH has not been
reported in detail in the literature. In this work we consider all
these models (LH, LHT-I, LHT-II and SLH) to perform a comparative
study.

Our work is organized as follows. In Sec. II we recapitulate the
models. In Sec. III we calculate the rate of $pp \to h \to
\gamma\gamma$ at the LHC in these models. Finally, we give
our conclusion in Sec. IV.

\section{little Higgs models}

\subsection{Littlest Higgs model (LH)}
The LH model \cite{lst,lhhantao} is based on a non-linear $\sigma$
model in the coset space of $SU(5)/SO(5)$ with additional local
gauge symmetry $[SU(2) \otimes U(1)]^2$. The vacuum expectation
value (VEV) of an $SU(5)$ symmetric tensor field breaks the $SU(5)$
to $SO(5)$ at the scale $f$. The top quark partner T-quark, heavy
gauge bosons ($W_H$, $Z_H$, $A_H$) and triplet scalar ($\Phi^{++}$,
$\Phi^+$, $\Phi^0$, $\Phi^P$) are respectively introduced  to cancel
the Higgs mass one-loop quadratic divergence contributed by the top
quark, gauge bosons and Higgs boson in SM.

The top quark and T-quark can give the dominant contributions to the
effective coupling $hgg$. Their Higgs couplings are given by
\beq
{\cal L}_t\simeq -\lambda_1 f \left[\frac{s_\Sigma}{\sqrt{2}}
\bar{u}_{L}u_R+ \frac{1+c_\Sigma}{2} \bar{U}_{L}u_R
\right]-\lambda_2f\bar{U}_LU_R+{\rm h.c.}, \label{lhtop}
\eeq
where
$c_\Sigma\equiv \cos\frac{\sqrt{2}(v+h)}{f}$ and $s_\Sigma\equiv
\sin\frac{\sqrt{2}(v+h)}{f}$, with $h$ and $v$ being the neutral
Higgs boson field and its VEV, respectively. After diagonalization
of the mass matrix in Eq. (\ref{lhtop}), we can get the mass
eigenstates $t$ and $T$ as well as their couplings with the Higgs
boson \cite{hrrhan},
\beq
 {\cal L}=
 -\frac{m_t}{v} y_{t} \bar{t}t h - \frac{m_T}{v} y_{_T}
 \bar{T}T h,
\eeq where \beq  m_T=\frac{m_tf}{s_t c_t v},\qquad y_{t} =
1+\frac{v^2}{f^2} \left[ - \frac{2}{3}+\frac{x}{2}-\frac{x^2}{4} +
c_t^2 s_t^2 \right],\qquad y_{_{T}} = -c_t^2s_t^2\frac{v^2}{f^2}.
\label{ytybigt} \eeq The parameter $x$ is a free parameter of the
Higgs sector proportional to the triplet VEV $v'$ and defined as $x
= \frac{4fv'}{v^2}$. The $c_t$ and $s_t$ are the mixing parameters
between $t$ and $T$, \beq r=\frac{\lambda_1}{\lambda_2},~~
c_t=\frac{1}{\sqrt{r^2+1}},~~ s_t=\frac{r}{\sqrt{1+r^2}}. \eeq In
addition to the Higgs couplings with charged fermions, the Higgs
couplings with the charged bosons also contribute to the effective
coupling $h\gamma\gamma$, which are given as
\begin{eqnarray}\label{lhvinter}
 {\cal L}&=&
 2\frac{m_{W}^2}{v} y_{_W} W^+ W^- h
 + 2\frac{m_{W_H}^2}{v} y_{_{W_H}} W^{+}_H W^{-}_H h \nonumber\\
&&
 -2\frac{m_{\Phi}^2}{v} y_{_{\Phi^+}} \Phi^+ \Phi^- h
 - 2\frac{m_{\Phi}^2}{v} y_{_{\Phi^{++}}} \Phi^{++}
  \Phi^{--} h,
\end{eqnarray}
where
\begin{eqnarray}
    \label{yi}
    \begin{array}{rclrcl}
    m_{W_H} &=& \frac{gf}{2sc},
                   \qquad &
    m_{\Phi}&=& \frac{\sqrt{2}m_h}{\sqrt{1-x^2}}\frac{f}{v},\\
    y_{_{W_L}}&=&1+ \frac{v^2}{f^2}\left[-\frac{1}{6}-\frac{1}{4}(c^2-s^2)^2\right],
                   \qquad &
    y_{_{W_H}}&=&- s^2c^2\frac{v^2}{f^2}, \\
    y_{\Phi^+}&=& \frac{v^2}{f^2}\left[-\frac{1}{3}+\frac{1}{4}x^2\right], \qquad &
    y_{\Phi^{++}}&=& \frac{v^2}{f^2}\ord(\frac{x^2}{16}\frac{v^2}{f^2},\frac{1}{16\pi^2}).
    \end{array}
\end{eqnarray}
The $c$ and $s$ are the mixing parameters in the gauge boson sector.
Since the $h\Phi^{++}\Phi^{--}$ coupling is very small, the
contributions of the doubly-charged scalar can be ignored. In the
littlest Higgs model, the relation between $G_F$ and $v$ is modified
from its SM form, which can induce \cite{hrrhan} \beq v \simeq
v_{SM}[1-\frac{v^2_{SM}}{f^2}(-\frac{5}{24}+\frac{1}{8}x^2)], \eeq
where $v_{SM}=246$ GeV is the SM Higgs VEV.

\subsection{Littlest Higgs models with T-parity (LHT)}

The LHT-I and LHT-II have the same kinetic term of $\Sigma$ field
where the T-parity can be naturally implemented, requiring that the
coupling constant of $SU(2)_1$ ($U(1)_1$) equals to that of
$SU(2)_2$ ($U(1)_2$). This will make the four mixing parameters in
gauge sector $c$, $s$ $(\equiv\sqrt{1-c})$, $c'$ and $s'$
$(\equiv\sqrt{1-c'})$ equal to $1/\sqrt{2}$, respectively. Under
T-parity, the SM bosons are T-even and the new bosons are T-odd.
Therefore, the coupling $H^{\dag}\phi H$ is forbidden, leading the
triplet VEV $v'=0$ and $x=0$. Since the correction of $W_H$ to the
relation between $G_F$ and $v$ is forbidden by T-parity, the Higgs
VEV $v$ is modified as \cite{lhtiyuan,higgslhtii} \beq v\simeq
v_{SM}(1+\frac{1}{12}\frac{v^2_{SM}}{f^2}). \eeq The Higgs couplings
with charged bosons of LHT-I and LHT-II can be obtained from the
Eq.(\ref{lhvinter}) and Eq. (\ref{yi}) by taking $c=s=1/\sqrt{2}$
and $x=0$.

For each SM quark (lepton), a heavy copy of mirror quark (lepton)
with T-odd quantum number is added in order to preserve the
T-parity. In the LHT-I \cite{lhti,lhtiyuan,lhtihubisz}, the T-parity
is simply implemented by adding the T-parity images for the original
top quark interaction to make the Lagrangian T-invariant, so that
the top quark partner canceling the one-loop quadratic divergence of
Higgs mass is still T-even. Inspired by the way that the quadratic
divergence given by top quark is canceled in the SLH, ref.
\cite{lhtii} takes an alternative implementation of T-parity in
LHT-II, where all new particles including the heavy top partner
responsible for canceling the SM one-loop quadratic divergence are
odd under T-parity.

In the LHT-I, the Higgs couplings with the heavy quarks are given by
\begin{eqnarray} {\cal L}_{\kappa}&\simeq& -\sqrt{2} \kappa f
\left[\frac{1+c_\xi}{2} \bar{u}_{L_-} u'_R
-\frac{1-c_\xi}{2}\bar{u}_{L_-}q_R
-\frac{s_\xi}{\sqrt{2}}\bar{u}_{L_-} \chi_R\right] \nonumber \\
&& - m_q \bar{q}_Lq_R - m_{\chi}\bar{\chi}_L\chi_R +{\rm
h.c.},\label{lhti-odd}
\end{eqnarray}\beq {\cal L}_t\simeq -\lambda_1 f \left[\frac{s_\Sigma}{\sqrt{2}}
\bar{u}_{L_+}u_R+ \frac{1+c_\Sigma}{2} \bar{U}_{L_+}u_R
\right]-\lambda_2f \bar{U}_{L_+}U_{R_+}+{\rm h.c.}, \label{lhti-t}
\eeq where $c_\xi \equiv \cos\frac{v+h}{\sqrt{2}f}$ and $s_\xi\equiv
\sin\frac{v+h}{\sqrt{2}f}$. After diagonalization of the mass matrix
in Eq. (\ref{lhti-odd}), we can get the T-odd mass eigenstates
$u_-$, $q$ and $\chi$. In fact, there are three generations of T-odd
particles, and we assume they are degenerate. The mass eigenstates
$t$ and $T$ can be obtained by mixing the interaction eigenstates in
Eq. (\ref{lhti-t}), and their Higgs couplings are the same to those
of LH with $x=0$.

In the LHT-II,
the Higgs couplings with the first two generations of heavy quarks
are given by \beq \label{lhtii-12} {\cal L}_{q}^{1,2}\simeq-\sqrt{2}
\kappa f \left[\frac{1+c_\xi}{2} \bar{u}_{L_-} u'_R
-\frac{1-c_\xi}{2}\bar{u}_{L_-}q_R
+\frac{s_\xi}{\sqrt{2}}\bar{u}_{L_+} \chi_R \right]- m_q
\bar{q}_Lq_R- m_{\chi}\bar{\chi}_L\chi_R+{\rm h.c.} .
\eeq
The mass
eigenstates of $u_-$, $q$ and $\chi$ and their Higgs couplings can
be obtained by the diagonalization of the mass matrix in Eq.
(\ref{lhtii-12}).

The Higgs couplings with the third generation of heavy quarks are
given by
\begin{eqnarray} {\cal L}_{q}^{3}
&\simeq&-\sqrt{2} \kappa f \left [\frac{1+c_\xi}{2} \bar{u}_{L_-}
u'_R -\frac{1-c_\xi}{2}\bar{u}_{L_-}q_R
-\frac{s_\xi}{\sqrt{2}}\bar{U}_{L_-}q_R
  -\frac{s_\xi}{\sqrt{2}}\bar{U}_{L_-}u'_R
  +\frac{s_\xi}{\sqrt{2}}\bar{u}_{L_+} \chi_{R}
  \right. \nonumber \\&&\left.+c_\xi\bar{\chi}_{L}\chi_{R} \right]- m_q
\bar{q}_Lq_R -\lambda f \left[s_\Sigma
\bar{u}_{L_+}u_{R_+}+\frac{1+c_\Sigma}{\sqrt{2}} \bar{U}_{L_-}
U_{R_-} \right]+{\rm h.c.}, \label{lhtii-3}
\end{eqnarray}
where $c_t$ is taken as $1/\sqrt{2}$. After diagonalization of the
mass matrix in Eq. (\ref{lhtii-3}), we can get the mass eigenstates
$t$, $T_-$, $u_-$, $q$ and $\chi$ as well as their Higgs couplings.

For the SM down-type quarks (leptons), the Higgs couplings of LHT-I
and LHT-II have two different cases \cite{lhtiyuan}
\begin{eqnarray}
\frac{g_{hd\bar{d}}}{g_{hd\bar{d}}^{\rm SM}}
&\simeq&1-\frac{1}{4}\frac{v_{SM}^2}{f^2}+\frac{7}{32}
\frac{v_{SM}^4}{f^4} ~~~~{\rm for~Case~A}, \label{Higgs-downA} \nonumber\\
&\simeq&1-\frac{5}{4}\frac{v_{SM}^2}{f^2}-\frac{17}{32}
  \frac{v_{SM}^4}{f^4} ~~~~{\rm for~Case~B}.\nonumber
\label{eq15}
\end{eqnarray}
The relation of down-type quark couplings also applies to the lepton
couplings.

\subsection{Simplest little Higgs model (SLH)}
The SLH \cite{sst} model is based on $[SU(3) \times U(1)_X]^2$
global symmetry. The gauge symmetry $SU(3) \times U(1)_X$ is broken
down to the SM electroweak gauge group by two copies of scalar
fields $\Phi_1$ and $\Phi_2$, which are triplets under the $SU(3)$
with aligned VEVs $f_1$ and $f_2$.

The gauged $SU(3)$ symmetry promotes the SM fermion doublets into
$SU(3)$ triplets. The Higgs couplings with the quarks are given by
\bea \label{tTmixing}
{\cal L}_t &\simeq&-f \lambda_2^t \left[
x_\lambda^t c_\beta t_1^{c'}(-s_1t'_L
   +c_1T'_L)+s_\beta t_2^{c'} (s_2 t'_L+ c_2 T'_L)\right]+h.c.,\,\\
   \label{dDmixing}
{\cal L}_{d} &\simeq&-f \lambda_2^{d} \left[ x_\lambda^{d} c_\beta
d_1^{c'}
  (s_1 d'_{L}+c_1 D'_{L})+s_\beta d_2^{c'} (-s_2 d'_{L}+c_2 D'_{L})\right]+h.c.,\,\\
\label{sDmixing} {\cal L}_{s} &\simeq&-f \lambda_2^{s} \left[
x_\lambda^{s} c_\beta s_1^{c'}
  (s_1 s'_{L}+c_1 S'_{L})+s_\beta s_2^{c'} (-s_2 s'_{L}+c_2 S'_{L})\right]+h.c.,\, ,
\eea where $f=\sqrt{f_1^2+f_2^2}$, $t_\beta\equiv
tan\beta=\frac{f_2}{f_1}$, $c_\beta=\frac{f_1}{f}$,
$s_\beta=\frac{f_2}{f}$, and
\bea
s_1\equiv \sin {t_\beta (h+v)\over
\sqrt{2}f},\ \ s_2\equiv \sin{(h+v) \over \sqrt{2}t_\beta f},\ \
s_3\equiv \sin{(h+v)(t_\beta^2+1)\over \sqrt{2}t_\beta f}.
 \eea
After diagonalization of the mass matrix in Eqs. (\ref{tTmixing}),
(\ref{dDmixing}) and (\ref{sDmixing}), we can get the mass
eigenstates $(t,~T)$, $(d,~D)$ and $(s,~S)$, which was performed
numerically in our analysis, and the relevant couplings with Higgs
boson can be obtained.

The Higgs coupling with the charged bosons is given by
\cite{slhvdefine},
\beq
 {\cal L}=
 2\frac{m_{W}^2}{v} y_{_W} W^+ W^- h+2\frac{m_{W'}^2}{v} y_{_{W'}} W^{'+} W^{'-} h,
 \eeq
where
\beq m^2_{W^{'+}} = \frac{g^2}{2}f^2, ~ y_{_W} \simeq
\frac{v}{v_{SM}} \left[ 1- \frac{v_{SM}^2}{
4f^2}\frac{\tbt^4-\tbt^2+1}{\tbt^2} + \frac{v_{SM}^4}{36
f^4}\frac{(\tbt^2-1)^2}{\tbt^2} \right],~ y_{_{W'}} \simeq
-\frac{v^2}{2f^2}.
\eeq

The Yukawa and gauge interactions break the global symmetry and then
provide a potential for the Higgs boson. However, the
Coleman-Weinberg potential alone is not sufficient since the
generated Higgs mass is too heavy. Therefore, one can introduce a
tree-level $\mu$ term which can partially cancel the Higgs mass,
\beq -\mu^2 (\Phi^\dagger_1 \Phi_2 + h.c.) = - 2 \mu^2 f^2 \sbt\cbt
\cos\left( \frac{\eta}{\sq \sbt\cbt f} \right)
 \cos \left(
 \frac{\sqrt{H^\dagger H}}{f \cbt\sbt}
\right).
\end{equation}
Where $\eta$ is a pseudo-scalar boson, whose mass is determined by
the parameter $\mu$.

 The Coleman-Weinberg potential involves the following
parameters
 \beq \label{para}
 f,~ x_\lambda^t,~ t_\beta,~\mu,~m_h,v.
\end{equation}
Due to the modification of the observed $W$-boson mass, $v$ is
defined as \cite{slhvdefine}
\beq \label{eq:v} v \simeq v_{SM}
\left[ 1+ \frac{v_{SM}^2}{12 f^2}\frac{\tbt^4-\tbt^2+1}{\tbt^2} -
\frac{v_{SM}^4}{180 f^4}\frac{\tbt^8-\tbt^6+\tbt^4-\tbt^2+1}{\tbt^4}
\right].
\end{equation}
Assuming that there are no large direct contributions to the
potential from physics at the cutoff, we can determine other
parameters in Eq. (\ref{para}) from $f$, $t_\beta$ and $m_h$ with
the definition of $v$ in Eq. (\ref{eq:v}).

\section{The di-photon $pp \to h \to \gamma\gamma$ signal at LHC}
\subsection{Calculations}
At the LHC the cross section of the single Higgs production via gluon
-gluon fusion can be given
\bea
\sigma(pp\to (gg\to h)X)\equiv
\sigma(gg\to h)&=&\tau_0\int_{\tau_0}^1
\frac{dx}{x}f_{g}(x,~\mu^2_F)f_{g}(\frac{\tau_0}{x},~\mu^2_F)\hat{\sigma}(gg\to
h),\nonumber\\
\hat{\sigma}(gg\to h)&=&\Gamma(h\to gg)\frac{\pi^2}{8m^3_h},
\label{gghhgg}
\eea
where $\tau_0=\frac{m_h^2}{s}$ with $\sqrt{s}$
being the center-of-mass energy of the LHC and $f_{g}(x,~\mu^2_F)$
is the parton distributions of gluon. The Eq. (\ref{gghhgg}) shows
that the $\sigma(gg\to h)$ has the strong correlation with decay
width $\Gamma(h\to gg)$.

Now we discuss the Higgs decays in little Higgs models. For the
tree-level decays $h\to XX$ where $XX$ denotes $WW$, $ZZ$ or the SM
fermion pairs, the little Higgs models give the correction via the
corresponding modified couplings
\beq \Gamma(h \to XX)= \Gamma(h \to
XX)_{SM}(g_{hXX}/g_{hXX}^{SM})^2.
\end{equation}
 $\Gamma(h \to
XX)_{SM}$ is the SM decay width, and $g_{hXX}$ and $g_{hXX}^{SM}$ are
the couplings of $hXX$ in the little Higgs models and SM,
respectively.

For a low Higgs mass, the loop-induced decay $h \to gg$ will be
important. The general expression for the effective coupling $hgg$
are shown in Appendix A. In the SM, the main contributions are from
the top quark loop, and the little Higgs models give the corrections
via the modified couplings $ht\bar{t}$. In addition, the decay width
of $h\to gg$ can be also corrected by the loops of heavy partner
quark $T$ quark in LH ($T,~D$ and $S$ in SLH) (new T-even and T-odd
quarks in LHT-I and LHT-II).

The general expression for the effective coupling $h\gamma\gamma$
are shown in Appendix A. In the SM,  the top quark loop and
$W$-boson loop give the main contributions to the decay $h\to
\gamma\gamma$. The little Higgs models give the corrections via the
modified couplings $ht\bar{t}$ and $hWW$. In addition to the loops
of the heavy quark mentioned in the decay $h\to gg$, the decay width
of $h\to \gamma\gamma$ can be also corrected by the loops of $W_H$,
$\Phi^+$, $\Phi^{++}$ in the LH, LHT-I and LHT-II ($W'$ in the SLH).
Note that in the lepton sector, LHT-I, LHT-II and SLH also predict
some neutral heavy neutrinos, which do not contribute to the
couplings of $h\gamma\gamma$ at the one-loop level. Although the
charged heavy leptons are predicted in LHT-I and LHT-II, they do not
have direct couplings with the Higgs boson.

In addition to the SM decay modes, the Higgs boson in the LHT-I,
LHT-II and SLH has some new important decay modes which are
kinematically allowed in the parameter space. In the LHT-I the
breaking scale $f$ may be as low as 500 GeV \cite{flht-i}, and the
constraint in LHT-II is expected to be even weaker \cite{lhtii}. For
a lower value of $f$, the lightest T-odd particle $A_H$ may have a
light mass, so that the decay $h\to A_H A_H$ can be open, whose
partial width is
\bea
\Gamma(h \to A_{H} A_{H}) & = & \frac{g_{hA_H
A_H}^{2} m_h^3}{128 \pi m_{A_H}^4}
  \sqrt{1-x_{A_H}}\left(1-x_{A_H}+\frac{3}{4}x_{A_H}^2\right),
\eea
where $x_{A_H}=4m_{A_H}^2/m_{h}^2$, and $g_{hA_H A_H}$ is the
coupling constants of $hA_HA_H$. However, in the LH the electroweak
precision data requires $f$ larger than a few TeV
\cite{cstrnotparity} and thus the decay $h\to A_HA_H$ is
kinematically forbidden.

In the SLH, the new decay modes are $h\to \eta\eta$ and $h \to
Z\eta$, whose partial widths are given by
\bea \label{eq:Gamma:new}
\Gm(h \to \eta\eta) &=&
 \frac{{\lambda'}^2}{8\pi}\frac{v^2}{m_h} \sqrt{1-x_\eta},\nonumber\\
\Gamma( h \to Z \eta) &=& \frac{m_h^3}{32 \pi f^2}
  \left( t_\beta - \frac{1}{t_\beta} \right)^2 \,
  \lambda^{3/2} \left(1, \frac{m_Z^2}{m_h^2}, \frac{m_\eta^2}{m_h^2}
 \right ),
\eea
where $x_\eta =4m_\eta^2/m_h^2$ and $\lambda (1,x,y) =
(1-x-y)^2 - 4 xy$.

\subsection{Numerical results and discussions}
In our calculations the SM input parameters involved are taken from
\cite{pdg}. For the SM decay channels, the relevant higher order QCD
and electroweak corrections are considered using the code Hdecay
\cite{hdecay}. We focus on a light SM-like Higgs boson, whose mass
is taken in the range of 110-140 GeV.

In the LH model the new free parameters are
$f,~c,~c',~c_t$ and $x$, where
\beq
0<c<1,~~~0<c'<1,~~~0<c_t<1,~~~0<x<1.
\eeq
Taking $f=1$ TeV, $f=2$
TeV and $f=4$ TeV, we scan over these parameters in the above
ranges and show the scatter plots. The parameter $c_t$ can control
the Higgs couplings with $t$, $T$ and $m_T$, which is involved in
the calculation of  $\Gamma(h\to t\bar{t})$, $\Gamma(h\to gg)$ and
$\Gamma(h\to \gamma\gamma)$. For a light Higgs boson, the decay mode
$h\to t\bar{t}$ is kinematically forbidden. For the $\Gamma(h\to
gg)$ and $\Gamma(h\to \gamma\gamma)$, the $c_t$ dependence of
top-quark loop and T-quark loop can cancel each other to a large
extent (see Eq. (\ref{ytybigt})). Therefore, the rate $\sigma(gg\to
h)\times BR(h\to \gamma\gamma)$ is not sensitive to $c_t$ for a
light Higgs boson.

\begin{figure}[tb]
 \epsfig{file=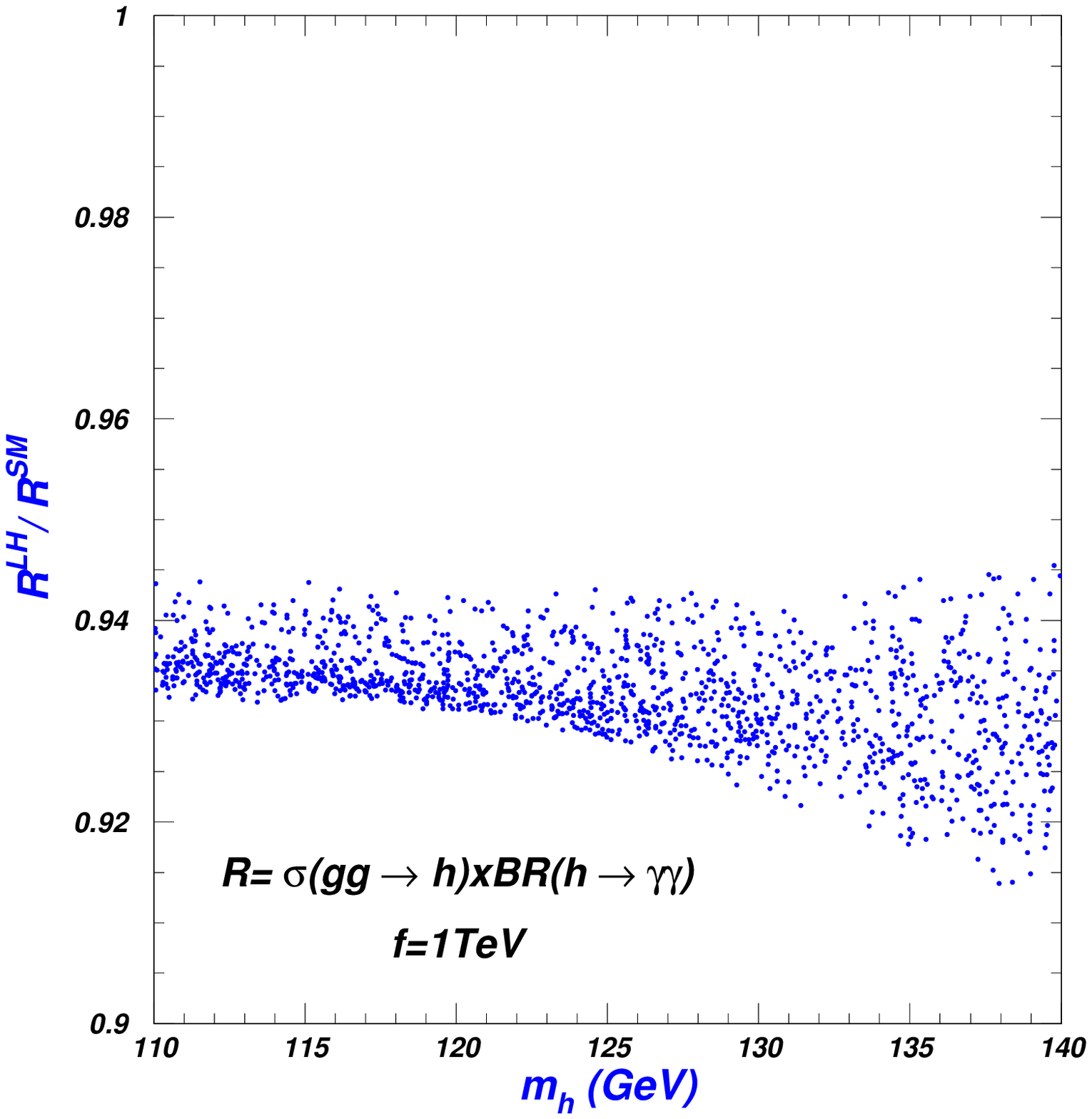,height=5.9cm}
 \epsfig{file=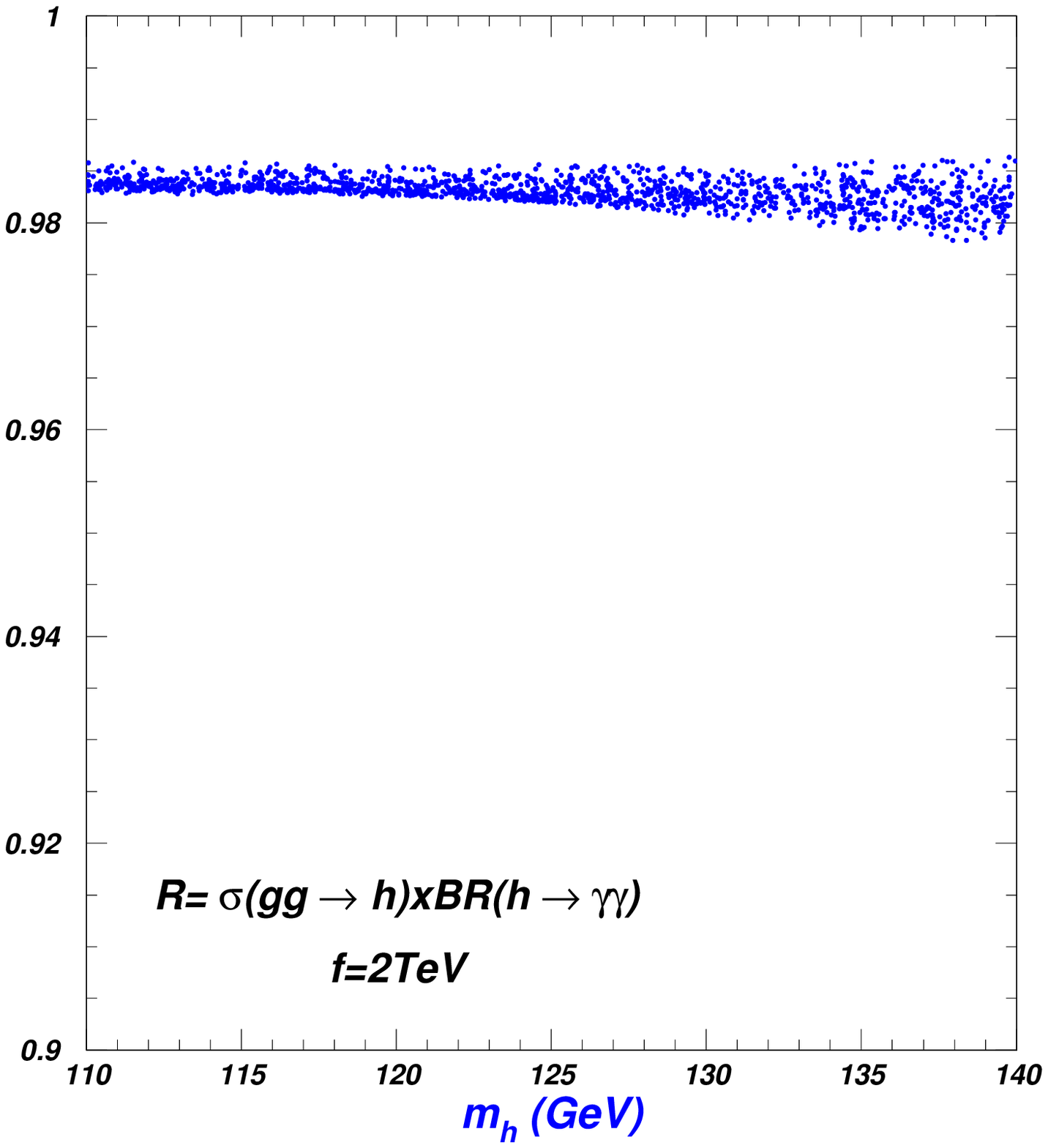,height=5.9cm}
 \epsfig{file=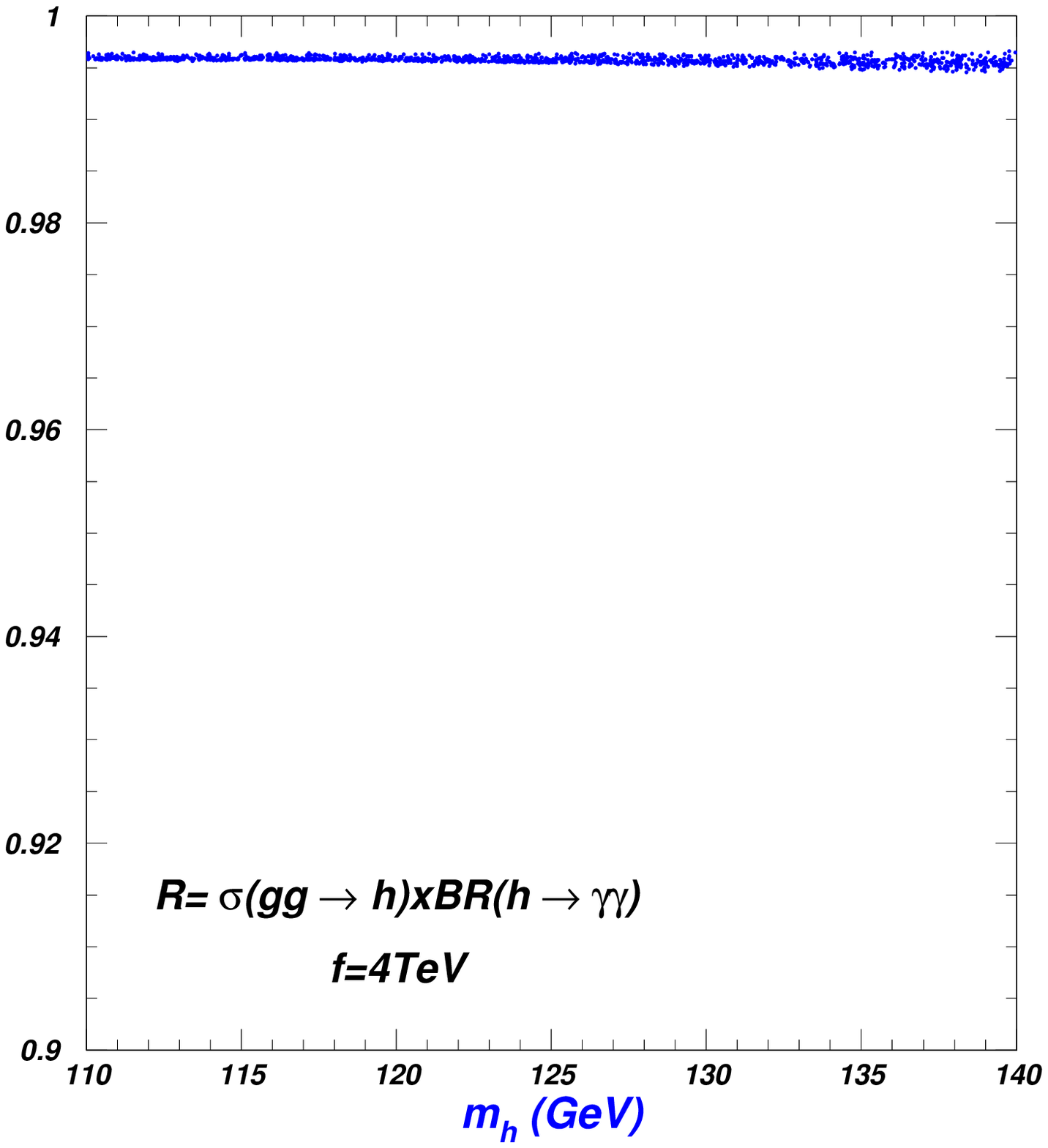,height=5.9cm}
\vspace{-1.0cm} \caption{Scatter plots for the rate $\sigma(pp\to
h)\times BR(h\to \gamma\gamma)$ at the LHC normalized to the SM
prediction in the LH model.} \label{figlh}
\end{figure}

\begin{figure}[htb]
 \epsfig{file=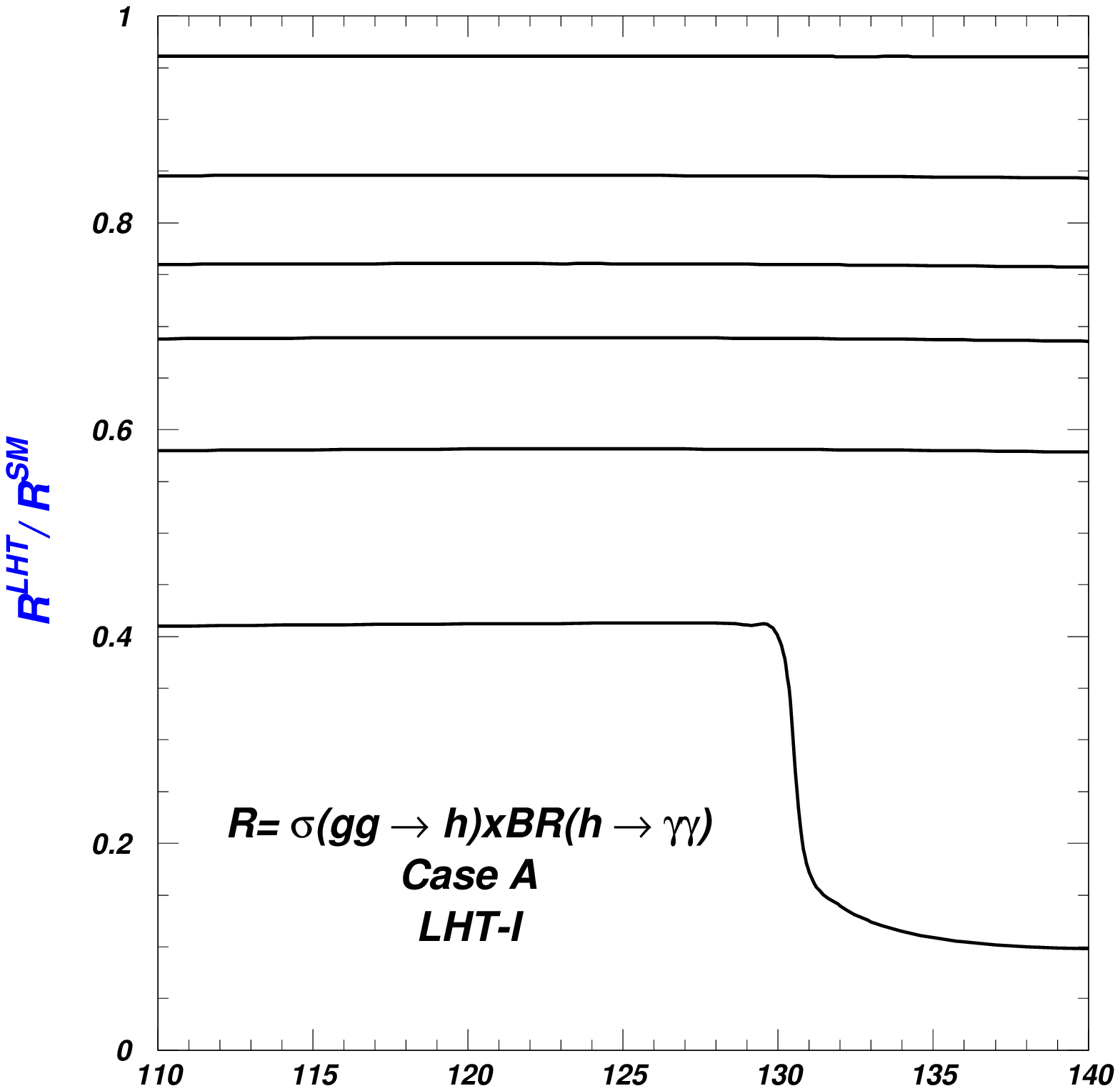,height=6.2cm}
 \epsfig{file=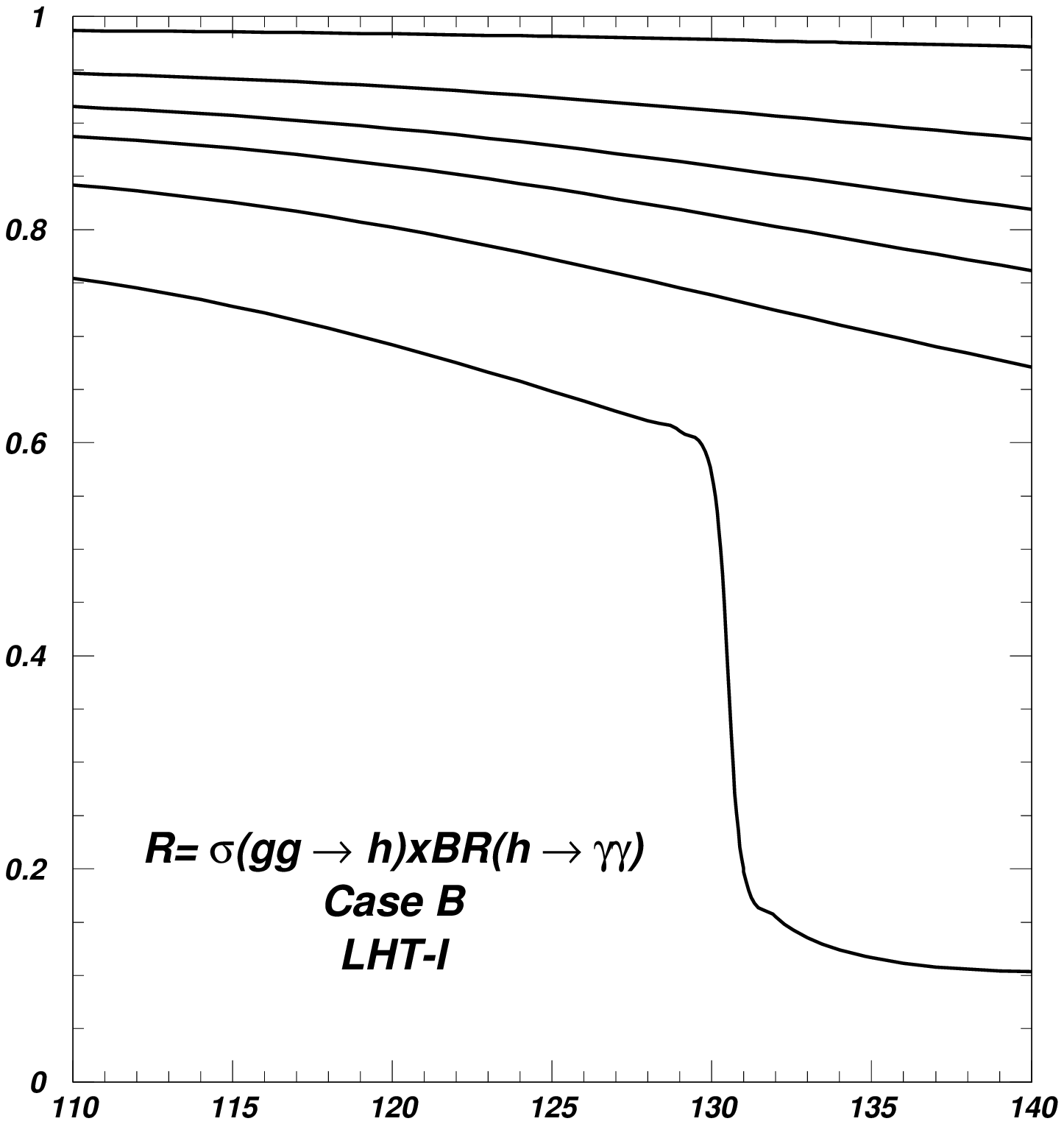,height=6.2cm}
 \epsfig{file=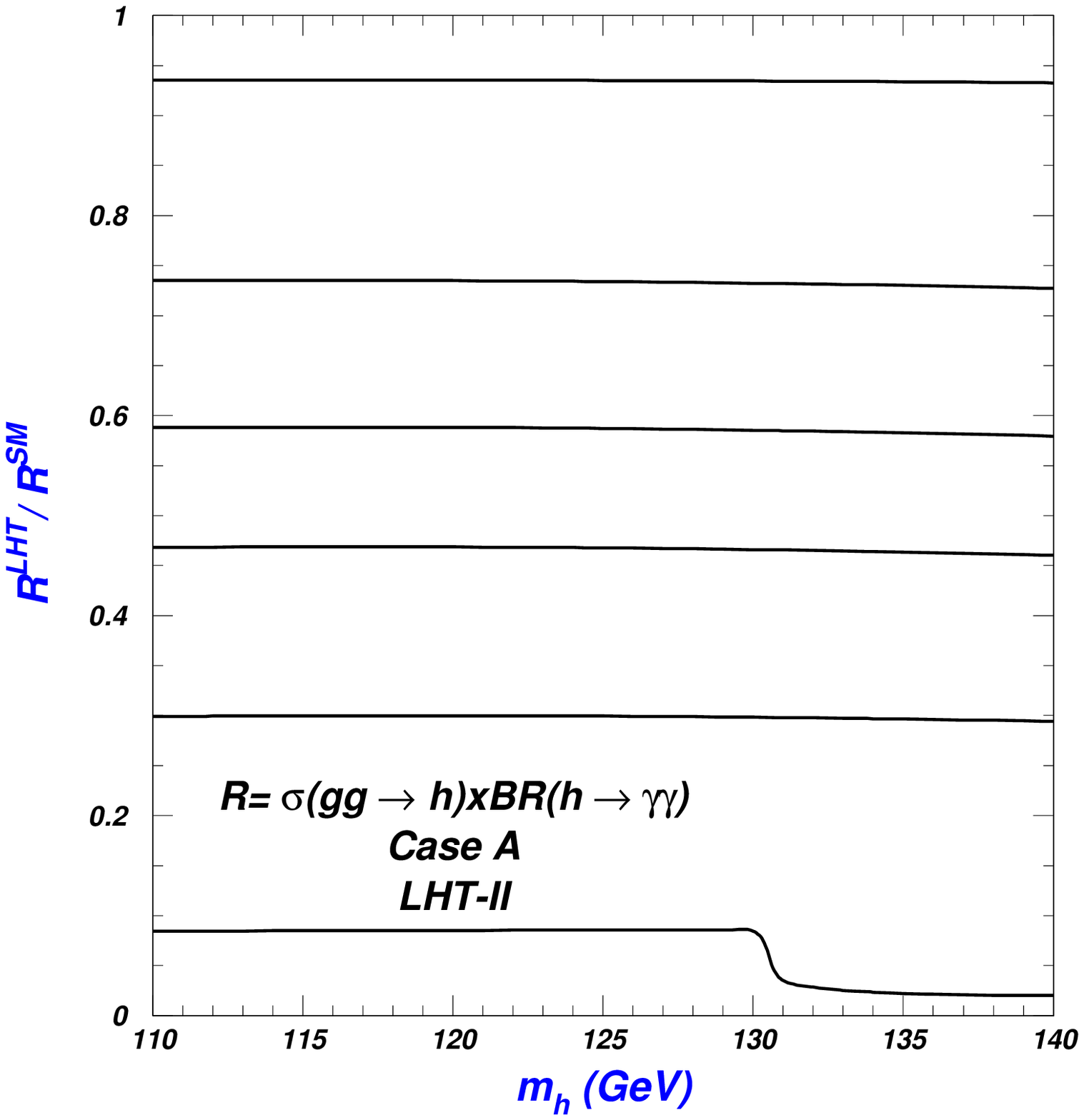,height=6.2cm}
 \epsfig{file=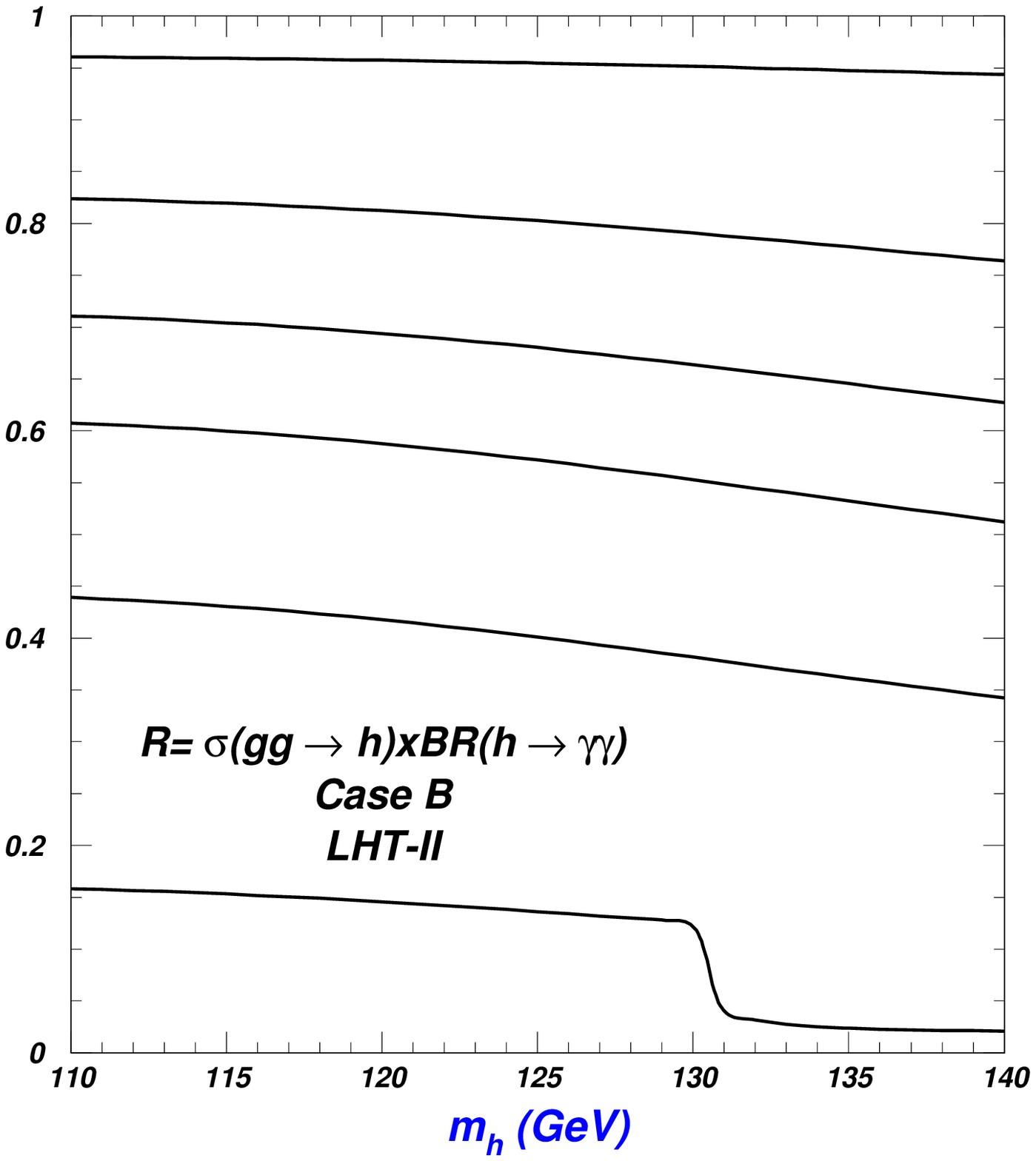,height=6.2cm}
\vspace{-.5cm} \caption{The rate $\sigma(pp\to h)\times BR(h\to
\gamma\gamma)$ at the LHC normalized to the SM prediction in the
LHT. The curves from bottom to top correspond to $f=$ 500 GeV, 600
GeV, 700 GeV, 800 GeV, 1 TeV and 2 TeV, respectively.}
\label{figlht}
\end{figure}

The rate $\sigma(pp\to h)\times BR(h\to \gamma\gamma)$ for the LH model
is shown in Fig. \ref{figlh} normalized to the SM prediction. We
can see that the LH model always suppresses the rate $\sigma(pp\to
h)\times BR(h\to \gamma\gamma)$, but the suppression can only reach
about $10\%$ for the small value $f$. As the increasing of $f$, the
magnitude becomes small, and the rate is not sensitive to the
parameters $c$, $c'$, $c_t$ and $x$. For example, for $f=4$ TeV, the
scatter plots are shown in line with the rate being around 99.6
percent of SM prediction.

In LHT-I and LHT-II, the parameters $c$, $c'$ and $x$ are fixed as $
c=c'=\frac{1}{\sqrt{2}}$ and $x=0.$ Similar to the LH model, the
result is not sensitive to $c_t$ in LHT-I and LHT-II. Taking
$c_t=1/\sqrt{2}$ can simplify the top quark Yukawa sector in the
LHT-II \cite{lhtii,higgslhtii}, and this choice is also favored by
the electroweak precision data \cite{flht-i}. The new heavy quarks
can contribute to the decay widths of $h\to gg$ and $h\to
\gamma\gamma$ via the loop, which are not sensitive to the actual
values of their masses as long as they are much larger than half of
the Higgs boson mass \cite{hrrhan}.

The rate $\sigma(pp\to h)\times BR(h\to \gamma\gamma)$ for LHT-I and
LHT-II is shown in Fig. \ref{figlht} normalized to the SM
prediction. We can see that LHT-I and LHT-II always suppress the
rate, and the suppression is much more sizable than that of LH. For
each model the rate in Case A is smaller than the rate in Case B
because the coupling $hb\bar{b}$ in Case A is less suppressed than
in Case B. Besides, we see that for $f=500$ GeV and $m_h$ in the
range of 130 GeV - 140 GeV, the rate in both models drops
drastically. The reason for such a severe suppression is that the
new decay mode $h\to A_HA_H$ is open and dominant in these parameter
space and thus the total decay width of Higgs boson becomes much
larger than the SM value.

\begin{figure}[htb]
 \epsfig{file=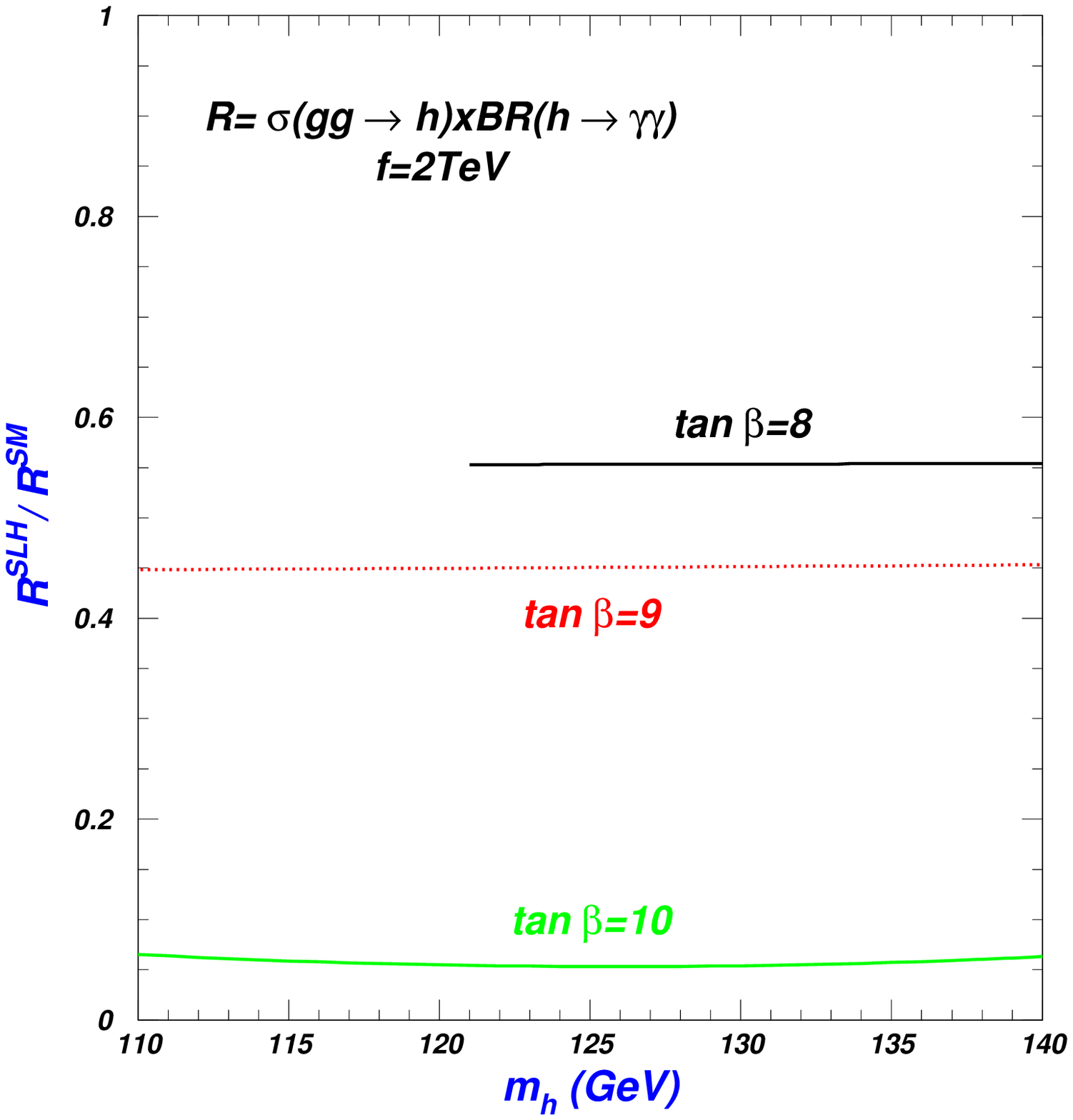,height=5.8cm}
 \epsfig{file=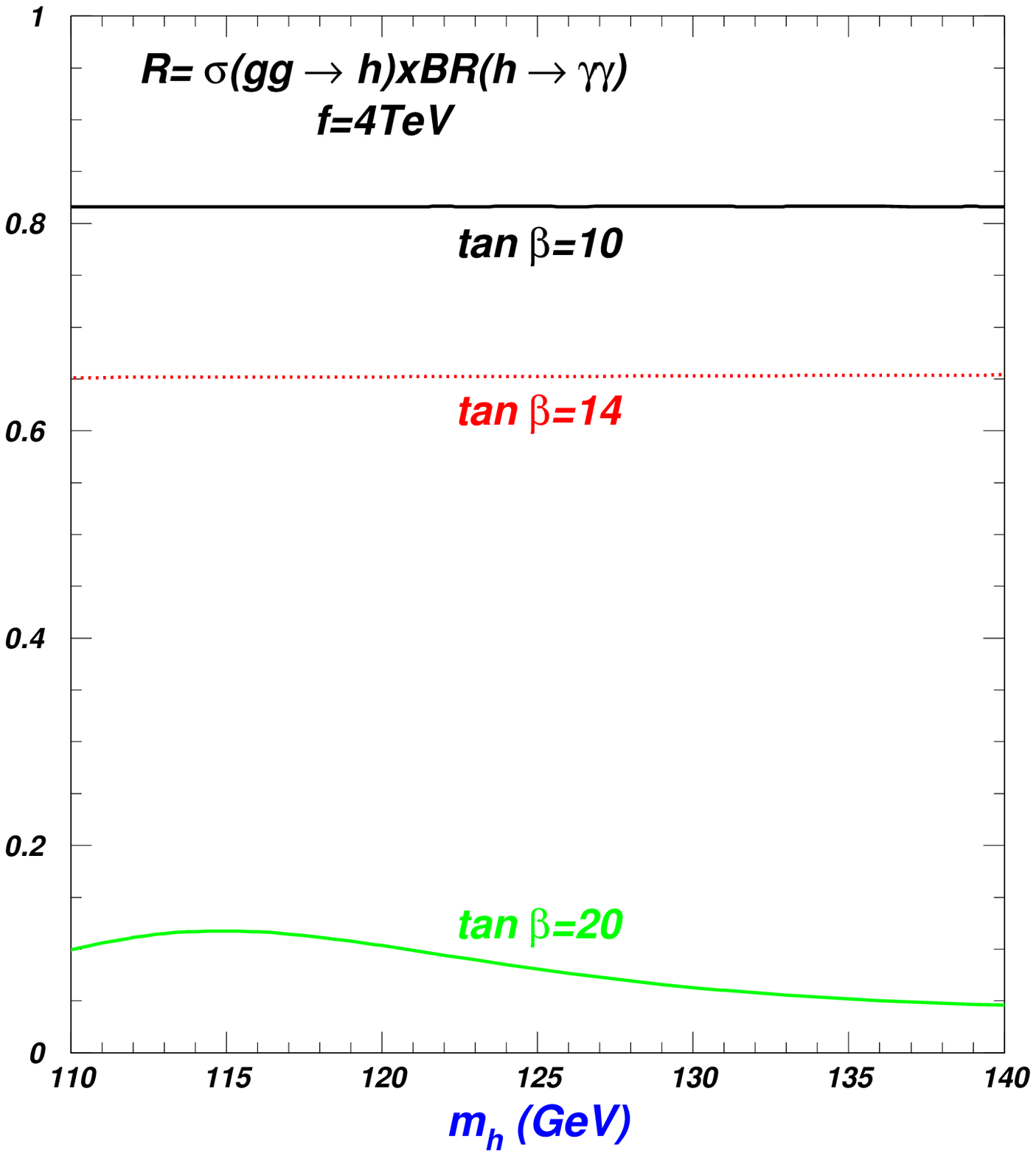,height=5.8cm}
 \epsfig{file=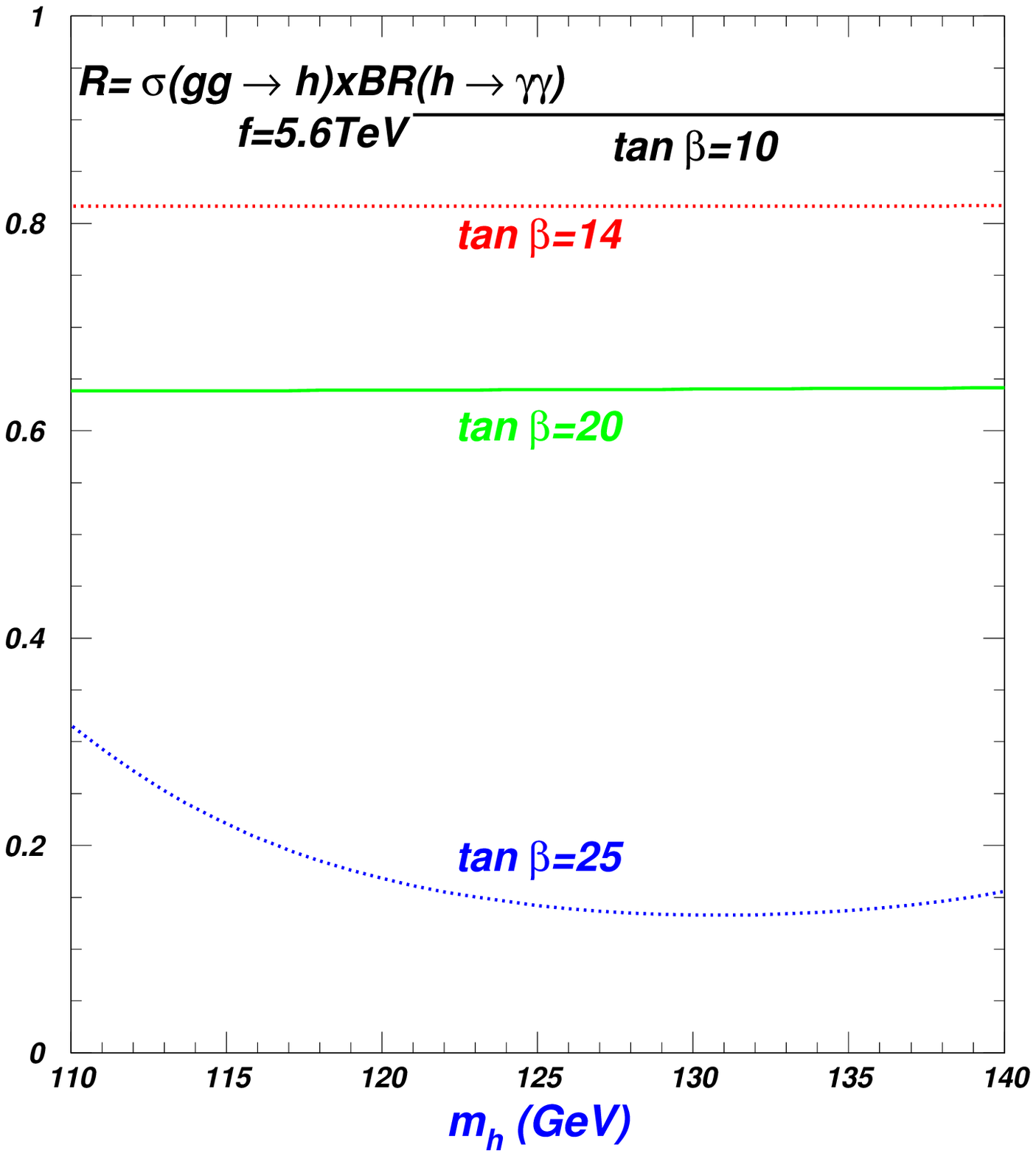,height=5.8cm}
\vspace{-.5cm} \caption{The rate $\sigma(pp \to h)\times BR(h\to
\gamma\gamma)$ at the LHC normalized to the SM prediction in SLH. The
incomplete lines for the small values of $\tan\beta$ show the lower
bounds of Higgs mass.}\label{figslh}
\end{figure}
In the SLH the new free parameters are $f, ~t_\beta,~
x_\lambda^{d}~(m_D)$ and $x_\lambda^{s}~(m_S)$. As shown above, the
parameters $x_\lambda^t,~\mu~(m_\eta)$ can be determined by $f$,
$t_\beta$, $m_h$ and $v$ with the assuming that the physics at the
cutoff does not give the large direct contributions to the
potential. Ref. \cite{sst} shows that the LEP-II data requires $f>2$
TeV, and ref. \cite{f4.5} gives a lower bound of $f>4.5$ TeV from
the oblique parameter $S$. A recent studies about $Z$ leptonic decay
and $e^+e^-\to \tau^+\tau^-\gamma$ process at the $Z$-pole show that
the scale $f$ should be respectively larger than 5.6 TeV and 5.4 TeV
\cite{f5.6}. Here, we assume the new flavor mixing matrices in
lepton and quark sectors are diagonal \cite{smoking,lpv}, so that
$f$ and $t_\beta$ are free from the experimental constraints of the
lepton and quark flavor violating processes. In addition, the
contributions to the electroweak precision data can be suppressed by
the large $t_\beta$ \cite{sst,newslh}. For the perturbation to be
valid, $t_{\beta}$ cannot be too large for a fixed $f$. If we
require $\ord(v_0^4/f^4)/\ord(v_0^2/f^2) < 0.1$ in the expansion of
$v$, $t_\beta$ should be below 10, 20, and 28 for $f=2$ TeV, 4 TeV,
and 5.6 TeV, respectively. The small masses of the $d$ quark and $s$
quark require that $x_\lambda^{d}$ and $x_\lambda^{s}$ are very
small, respectively, so there is almost no mixing between the SM
down-type quarks and their heavy partners, and the results are not
sensitive to them. We take $x_\lambda^{d}=1.1\times 10^{-4}$
($x_\lambda^{s}=2.1\times 10^{-3}$), which can make the masses of
$D$ and $S$ in the range of 0.5-2 TeV with other parameters fixed as
in our calculation.

The rate $\sigma(pp\to h)\times BR(h\to \gamma\gamma)$ for the SLH
at the LHC is shown in Fig. \ref{figslh}. We can
see that the SLH always suppresses the rate, and the suppression is
more sizable for a large $t_\beta$. When $t_\beta$ is large enough,
such as $t_\beta=10$ for $f=2$ TeV ($t_\beta=20$ for $f=4$ TeV or
$t_\beta=25$ for $f=5.6$ TeV), the new mode $h\to \eta\eta$ is open
and dominant, which can  further suppress the rate (the suppression
can be up to $90\%$).

\begin{figure}[htb]
 \epsfig{file=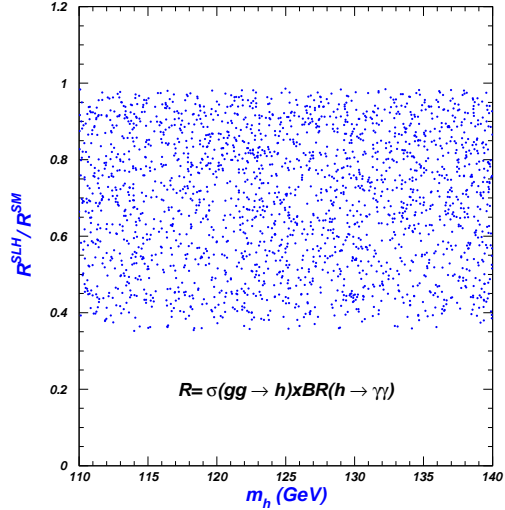,height=6.8cm}
\vspace{-.5cm} \caption{Scatter plots for the rate $\sigma(pp \to
h)\times BR(h\to \gamma\gamma)$ normalized to the SM prediction in
SLH. }\label{figslh2}
\end{figure}
If the ultraviolet completion of the theory can give the sizable
contributions to the Coleman-Weinberg potential, the correlation of
the parameters $x_\lambda^t,~\mu~(m_\eta),~f, ~t_\beta,~m_h$ and $v$
can be loosened greatly. In Fig. \ref{figslh2}, we scan the
following parameter space,
\begin{eqnarray}
&& 1~TeV < f < 6~TeV,~0.5~TeV< m_T <2~TeV, \nonumber \\
&& 0.5~TeV< m_D~(m_S)<2~TeV,~1<t_\beta<30,
\end{eqnarray}
where the parameter $x_\lambda^t$ is
replaced with $m_T$ and the bound $\ord(v_0^4/f^4)/\ord(v_0^2/f^2) <
0.1$ is still valid. To avoid that the rate is suppressed by the new
decay modes $h\to \eta\eta$ and $h\to \eta Z $, we take $m_\eta >2
m_h$, so that the result is independent of the parameter $m_\eta$
($\mu$). Fig. \ref{figslh2} shows that, compared to the SM
prediction, the SLH still suppresses the rate $\sigma(pp\to h)\times
BR(h\to \gamma\gamma)$ in more general parameter space.

\begin{table}[t]
\caption{The value of $y_{_{f_i}}$ corresponding to the quark $f_i$
and $Br(h\to \gamma\gamma)$ normalized to SM prediction in these
little Higgs models, respectively. The parameters are fixed as
$c=c'=c_t=\frac{1}{\sqrt{2}}$ and $x=0$ in the LH,  and
$t_\beta$=10, $m_T$=450 (798) GeV, $m_D$=548 (1039) GeV, $m_S$=597
(1132) GeV and $m_{\eta}$=42.7 (179) GeV for $f$=2 (4) TeV in the
SLH.}
\begin{center}
\begin{tabular}{|ll|c|c|c|c|c|}
\hline
      &$m_h$=120~GeV&~t-quark~&~t-quark partner~&~other heavy quarks~
        &$~\sum y_{_{f_i}}~$&$~\frac{Br(h\to\gamma\gamma)}{Br(h\to\gamma\gamma)_{SM}}~$\\
\hline LH & $f$=1 TeV   &0.974 &-0.016 &- &0.958 &1.040\\
          &$f$=2 TeV & 0.994& -0.004& -&0.990 & 1.009\\

\hline LHT-I & $f$=0.5 TeV   &0.895 &-0.080 &-0.195 &0.620 &1.094\\
 Case A      &$f$=1 TeV & 0.975& -0.016& -0.046&0.913 & 1.024\\
\hline LHT-I & $f$=0.5 TeV   &0.895 &-0.080 &-0.195 &0.620 &1.836\\
 Case B      &$f$=1 TeV & 0.975& -0.016& -0.046&0.913 & 1.130\\
\hline LHT-II & $f$=0.5 TeV   &0.745 &-0.338 &-0.164 &0.243 &1.381\\
 Case A      &$f$=1 TeV & 0.940& -0.084&-0.034&0.822 & 1.089\\
\hline LHT-II & $f$=0.5 TeV   &0.745 &-0.338 &-0.164 &0.243 &2.36\\
 Case B      &$f$=1 TeV & 0.940& -0.084&-0.034&0.822 & 1.204\\
\hline SLH & $f$=2 TeV   &0.776 &-0.117 &0.000 &0.659 &0.155\\
  &$f$=4 TeV & 0.978& -0.047& 0.000&0.931 & 0.998\\
\hline
\end{tabular}
\end{center}
\label{rmh120}
\end{table}

From our above results we see that compared to the SM prediction the
rate $\sigma(pp\to h)\times BR(h\to \gamma\gamma)$ is always
suppressed in these typical little Higgs models. Now we analyze such
a suppression in detail. Eq. (\ref{tau05}) and Eq. (\ref{limit})
show that the effective coupling $hgg$ is not sensitive to the heavy
quark masses as long as they are much larger than half of the Higgs
boson mass. Therefore, according to the Eq. (\ref{hgg}) and Eq.
(\ref{I05}), the effective coupling $hgg$ is approximately
proportional to $(\sum \frac{y_{_{f_i}}}{v})^2$, where $y_{_{f_i}}$
is defined in Eq. (\ref{rrinteri}). Table 1 shows the value of
$y_{_{f_i}}$ corresponding to the quark $f_i$ and $Br(h\to
\gamma\gamma)$ normalized to SM prediction in these little Higgs
models, respectively.

Because of the sizable suppression of the coupling $hb\bar{b}$,
$Br(h\to\gamma\gamma)$ is generally not suppressed, unless the new
decay mode is open and dominant, as shown for the SLH with $f=$ 2
TeV in Table 1 (the new mode is $h\to \eta\eta$). In these little
Higgs models, $\sum y_{_{f_i}}$ can be respectively less than 1,
which shows that the $\sigma(pp\to h)$ is suppressed compared the SM
prediction. There are some common reasons for these models: (i) All
the models are based on the non-linear sigma models, the Yukawa
coupling $ht\bar{t}$ is suppressed with the expansion of the
non-linear sigma fields. (ii) The top quark partner cancels the
quadratic divergence of Higgs mass contributed by top quark, which
will induce that the Yukawa couplings of top quark and its partner
have the opposite sign.

The forthcoming measurement of the di-photon signal at the LHC will
allow for a probe of these little Higgs models. For example, if the
signal rate is found to be above the SM prediction, these little
Higgs models will be immediately disfavored. If the signal rate is
found to be much lower than the SM prediction, then the SLH and LHT
will be favored. However, due to the free parameters involved in the
signal rate for each model, it will be hard for the LHC to clearly
discriminate these different little Higgs models. For the precision
test of different models, the ILC collider is necessary \cite{lei}.

\section{Conclusion}
We performed a comparative study for the LHC di-photon signal by
considering four different little Higgs models, namely the LH,
LHT-I, LHT-II and SLH. We obtained the following observations: (i)
Compared with the SM prediction, the di-photon signal rate is always
suppressed in these models; (ii) The suppression extent is different
in different models, which is below $10\%$ in the LH but can reach
$90\%$ in the LHT-I, LHT-II and SLH, especially in the parameter
space with new decay modes ($h\to \eta\eta$ for the SLH and $h\to
A_H A_H$ for the LHT-I and LHT-II) are open and dominant. Therefore,
discovering the light Higgs predicted by these little Higgs models
through the di-photon channel at the LHC will be more difficult than
discovering the SM Higgs boson.

\section*{Acknowledgment}
 JMY thanks JSPS for the invitation
fellowship (S-11028) and the particle physics group of
Tohoku University for their hospitality.
This work was supported in part by the National Natural Science
Foundation of China (NNSFC) under grant Nos. 11005089, 10821504,
10725526 and 10635030.

\appendix
\section{The effective couplings of
              Higgs-photon-photon and Higgs-gluon-gluon}
The effective Higgs-photon-photon coupling can be written as
\cite{hrrhan,higgshunter} \beq {\cal
L}^{eff}_{h\gamma\gamma}=-\frac{\alpha}{8\pi
v}IF_{\mu\nu}F^{\mu\nu}h, \label{hrr} \eeq where $F^{\mu\nu}$ is the
electromagnetic field strength tensor. With the Higgs boson
couplings to the charged fermion $f_i$, vector boson $V_i$ and
scalar $S_i$ given by \beq
 {\cal L}=\sum_{f_i}-\frac{m_{f_i}}{v}y_{_{f_i}}\bar{f_i}f_i h+ \sum_{V_i}2
 \frac{m_{_{V_i}}^2}{v} y_{_{V_i}} V_i V_i h
   +\sum_{S_i}- 2 \frac{m_{_{S_i}}^2}{v} y_{_{S_i}} S_i S_i h,
\label{rrinteri} \eeq the factor $I$ in Eq. (\ref{hrr}) can be
written as \beq I=\sum_{f_i} Q_{f_i}^2 N_{cf_i}~y_{_{f_i}}
I_{\frac{1}{2}}(\tau_{f_i})+\sum_{V_i} Q_{V_i}^2
~y_{V_i}I_1(\tau_{V_i})+\sum_{S_i} Q_{S_i}^2
~y_{S_i}I_0(\tau_{S_i}),
\eeq
where $Q_X$ ($X$ denotes $f_i$,
$V_i$ and $S_i$) is the electric charge for a particle $X$ running
in the loop, and $N_{cf_i}$ is the color factor for $f_i$. The
dimensionless loop factors are
\begin{eqnarray}\label{tau05}
I_{\frac{1}{2}}(\tau_{f_i})& =& -2\tau_{f_i} [1 + (1-\tau_{f_i})f(\tau_{f_i})], \\
I_1(\tau_{V_i}) &=& 2 + 3 \tau_{V_i} + 3\tau_{V_i}(2-\tau_{V_i}) f(\tau_{V_i}),\\
I_0(\tau_{S_i}) &=& \tau_{S_i} [1 - \tau_{S_i} f(\tau_{S_i})],
\end{eqnarray}
where $\tau_{_X} =4m_X^2/m_h^2$ and
\begin{equation}
    f(\tau_X) = \left\{ \begin{array}{lr}
        [\sin^{-1}(1/\sqrt{\tau_{X}})]^2, & \tau_{X} \geq 1 \\
        -\frac{1}{4} [\ln(\eta_+/\eta_-) - i \pi]^2, & \, \tau_{X} <
        1
        \end{array}  \right.\label{hggf12}
\end{equation}
with $\eta_{\pm}=1\pm\sqrt{1-\tau_X}$.  When the masses of particles
in the loops are much larger than half of the Higgs boson mass, we
can get
\begin{equation}\label{limit}
I_{\frac{1}{2}}(\tau_{f_i}) \simeq  -4/3, \qquad
I_1(\tau_{V_i}) \simeq  7, \qquad I_0(\tau_{S_i}) \simeq -1/3.
\end{equation}
The effective Higgs-gluon-gluon coupling can be written as
\cite{hrrhan,higgshunter}
\beq {\cal L}^{eff}_{hgg}=-\frac{\alpha_s}{12\pi v}I_{hgg}
G_{\mu\nu}^{\alpha}G^{\mu\nu}_{\alpha}h, \label{hgg}
\eeq
where
$G_{\mu\nu}^{\alpha}=\partial_{\mu} g_{\nu}^{\alpha}-\partial_{\nu}
g_{\mu}^{\alpha}$ and the factor $I_{hgg}$ from the contributions of
quarks running in the loops is given by
\beq
I_{hgg}=\sum_{q_i} \frac{3}{4}y_{q_i} I_{\frac{1}{2}}(\tau_{q_i}),
\label{I05}
\eeq
with $\tau_{q_i}=4m_{q_i}^2/m_h^2$.

Once the interactions in Eq. (\ref{rrinteri}) are given, we can
obtain the effective $h\gamma\gamma$ and $hgg$ couplings from the
above formulas. The relevant Higgs interactions in the LH, LHT-I and
LHT-II and SLH are listed in the Sec. II. Here the Higgs
interactions with the light fermions are not given since their
contributions can be ignored.

\end{document}